\documentclass[sigconf]{acmart}

\usepackage{graphicx}
\usepackage{subcaption}    
\usepackage{booktabs}
\usepackage[table]{xcolor} 
\usepackage{tabularx}
\usepackage[table,xcdraw]{xcolor}
\usepackage{pifont}

\newcommand{\cmark}{\textcolor{green!40!black}{\ding{51}}} 
\newcommand{\xmark}{\textcolor{red}{\ding{55}}}   

\usepackage{booktabs}

\AtBeginDocument{%
  }

\acmYear{2026}\copyrightyear{2026}
\setcopyright{cc}
\setcctype[4.0]{by}
\acmConference[ASIA CCS '26]{ACM Asia Conference on Computer and Communications Security}{June 1--5, 2026}{Bangalore, India}
\acmBooktitle{ACM Asia Conference on Computer and Communications Security (ASIA CCS '26), June 1--5, 2026, Bangalore, India}
\acmDOI{10.1145/3779208.3805984}
\acmISBN{979-8-4007-2356-8/26/06}

\begin{document}

\title{\textbf{\texttt{DECKER}}: Domain-invariant Embedding for Cross-Keyboard Extraction and Recognition}

\author{Bikrant Bikram Pratap Maurya}
\authornote{Both authors contributed equally to this research as First Authors.}
\affiliation{
  \institution{IIIT-Delhi}
  \country{India}}

\email{bikrant24116@iiitd.ac.in}

\author{Nitin Choudhury}
\authornotemark[1]
\authornote{Corresponding Author.}
\affiliation{
  \institution{IIIT-Delhi}
  \country{India}
}
\email{nitinc@iiitd.ac.in}

\author{Daksh Agarwal}
\affiliation{
  \institution{Guru Gobind Singh Indraprastha University, Delhi}
  \country{India}
  }

\email{dakshagrawal76@gmail.com}

\author{Arun Balaji Buduru}
\authornote{Primary Advisor.}
\affiliation{  
  \institution{IIIT-Delhi}
  \country{India}
}

\email{arunb@iiitd.ac.in}

\begin{abstract}

Acoustic side-channel attacks (ASCA) on keyboards pose a significant security risk, as keystrokes can be inferred from the typing acoustics, revealing sensitive information on laptops. Prior studies on ASCA are limited due to dataset constraints, which include a smaller number of users, keyboards, and environments. This limits the exploration of the potential of this attack vector across different keyboards, users, microphones, and ambient noise conditions.

To this end, we introduce \textbf{\texttt{HEAR}}, a novel dataset designed to study ASCA along three axes: keyboard generalization, noise adaptation, and user bias. \textbf{\texttt{HEAR}} contains recordings from 53 participants using 37 different laptop keyboards, collected in three realistic settings: (i) captured via an external microphone, and (ii) captured through the user's own device microphone over no network channel noise, and (iii) captured while in a network streaming platform over VoIP. This enables controlled benchmarks for cross-user, cross-keyboard, and cross-environment evaluations in a wider scope.

On \textbf{\texttt{HEAR}}, we establish a new ASCA benchmark that spans conventional features and pre-trained representations from both raw audio and spectrograms, in unimodal and multimodal settings. We further propose \textbf{\texttt{DECKER}}, a domain-invariant keystroke inference framework that works in a four-stage generalization strategy: (i) \emph{Keyboard Signature Normalization} to reduce device coloration in the waveform domain, (ii) \emph{domain-adversarial disentanglement} to suppress keyboard identity, (iii) \emph{supervised cross-keyboard contrastive alignment} to enforce key consistency across devices, and (iv) \emph{Acoustic Style Randomization} to synthesize unseen keyboard responses. To explore the potential depth of the attack, we further investigate a sentence-level inference attack that leverages a large language model (LLM) layer as a post-processing layer, refining the predicted keystroke sequences using linguistic context.

Empirical results on \textbf{\texttt{HEAR}} show that \textbf{\texttt{DECKER}} improves keystroke identification over strong unimodal and multimodal baselines, especially in cross-keyboard and cross-user settings, and that the language-model rectification further boosts sequence recovery. These findings highlight that ASCA remains effective under diverse users, devices, and noisy environments, underscoring its practical security risk.

\end{abstract}

\ccsdesc[500]{Security and privacy}
\ccsdesc[500]{Computing methodologies~Machine learning}
\ccsdesc[300]{Computing methodologies~Artificial intelligence}
\ccsdesc[100]{Hardware}

\keywords{
Acoustic Side Channel Attack,
ASCA,
Keystroke Inference,
Deep Learning,
Domain Invariance,
Cross-Keyboard Generalization,
ECAPA-TDNN,
Contrastive Learning,
Adversarial Representation Learning,
Audio Processing,
Security and Privacy
}

\received{}
\received[revised]{}
\received[accepted]{}

\maketitle


\section{Introduction}

An Acoustic Side-Channel Attack (ASCA) on a keyboard relies on the assumption that each keystroke produces a characteristic sound that can serve as a fingerprint for that key on a given keyboard. Prior work has empirically validated this assumption across multiple setups and threat models~\cite{asonov2004keyboard, harrison_practical_2023, zhou_patternlistener_2018, yu_indirect_2021, wang_accurate_2016, sabra_zoom_2020, compagno_dont_2017, das_you_2014, halevi_closer_2012, liu_when_2019, lu_keylistener_2019, martinasek_acoustic_2015}.

Laptops are now the primary platform for both personal and professional computing. Their portability allows them to be used in homes, libraries, caf\'es, shared workspaces, classrooms, and online meetings, across both private and public settings. In all these contexts, users routinely type sensitive information, such as passwords, PINs, confidential documents, and private messages, on their on-device physical keyboards. This convenience comes with a security cost: keystrokes can leak information through acoustic side channels. In practice, keystroke sounds can be passively captured not only by the nearby smartphones, concealed IoT devices, or other ambient microphones, but also from the laptop's own microphone. As microphones become pervasive in modern environments, the attack surface for ASCA expands accordingly, turning everyday laptop use into a potential threat to user privacy.

Currently, ASCA is particularly concerning as it is passive, covert, and deployable in ordinary social settings. A single device in a common space can silently record keystrokes from multiple users, independent of network boundaries or encryption. Further, typed inputs are rarely random in nature; they follow linguistic or structural patterns that adversaries can exploit. The emergence of large language models (LLMs)~\cite{radford2019gpt2, raffel2020t5, touvron2023llama} amplifies this threat significantly, making recovery of coherent text even from noisy or partially incorrect keystroke predictions plausible.

Previous research on keyboard ASCA has mostly been evaluated under narrow and idealized conditions. This includes setups that involve limited demographic variation among participants, low and stable noise levels, fixed device geometry, uniform keyboard switch mechanisms, and consistent typing habits~\cite{asonov2004keyboard,fadake_beyond_2023,harrison_practical_2023}. Recent work, such as RefleXnoop and Heimdall, partially relaxes the geometric assumptions by developing systems which demonstrate that multipath reflections can enable non-line-of-sight (NLoS) keystroke recovery~\cite{wang_reflexnoop_2024}.

On the modeling side, ASCA pipelines have progressed from early signal-processing classifiers to modern probabilistic and deep learning approaches. Initial work relied on temporal and spectral features such as FFT coefficients, MFCCs, and power or energy profiles, combined with statistical classifiers~\cite{song2001timing,zhuang2009keyboard}. Subsequent systems incorporated hidden Markov models (HMMs)~\cite{foo2010timing} and neural architectures~\cite{asonov2004keyboard,harrison_practical_2023} to better capture temporal structure and variability in keystroke acoustics. Recent efforts adopt convolutional and transformer-based models on audio spectrograms~\cite{ayati2025making} and show that LLMs can correct noisy keystroke predictions when linguistic context is available~\cite{ayati2025making}.

\textit{Despite these advances, most ASCA studies still assume controlled environments and single-device training regimes. They rarely account for the variability of real-world settings, where posture changes, ambient noise, reflective surfaces, and nearby devices introduce substantial acoustic distortion, and where users type on a wide range of laptop keyboards. As a result, the real-world generalizability and effectiveness of current ASCA techniques remain underexplored. Additionally, user biases, such as gender, age group, and hand dominance, play significant roles in ASCA, influencing factors including typing speed, key pressure, and typing errors. These gaps obscure realistic assessments of ASCA feasibility. Importantly, no prior work systematically evaluates ASCA across demographic diversity or examines user awareness of acoustic leakage threats. This highlights the need for a dataset and evaluation framework that spans diverse users, keyboards, and environments under realistic threat models, while explicitly enabling analysis of user bias and human factors in acoustic side-channel attacks.}

Motivated by these gaps, we first propose \textbf{\texttt{HEAR}}, a novel dataset designed for ASCA evaluation under a realistic threat model. Unlike existing ASCA datasets, which are highly constrained—typically single-keyboard, single-environment, and demographically narrow (often limited in gender distribution, typing backgrounds, and laptop usage profiles)~\cite{wang_reflexnoop_2024, harrison_practical_2023}, \textbf{\texttt{HEAR}} allows evaluation of ASCA under user biases, keyboard, and environment diversity.

On \textbf{\texttt{HEAR}}, we establish a new ASCA benchmark using strong audio and vision pre-trained models (PTMs) on both raw waveforms and spectrograms in unimodal and multimodal settings. To improve generalization across keyboards, environments, and users, we design \textbf{\texttt{DECKER}} as an ASCA training framework that unifies five key components: \textit{Keyboard Signature Normalization (KSN)} to reduce device-specific artifacts at the waveform level, \textit{domain-adversarial disentanglement} via gradient reversal~\cite{ganin2015unsupervised}, \textit{cross-keyboard supervised contrastive alignment}~\cite{khosla2020supervised}, \textit{Acoustic Style Randomization (ASR)} to simulate unseen keyboard acoustics~\cite{park2019specaugment}, and \textit{LLM-assisted decoding} with constrained beam search to reconstruct coherent text or passwords~\cite{radford2019gpt2,raffel2020t5}. Together, these components yield embeddings that are \emph{key-discriminative}, \emph{keyboard-invariant}, and \emph{linguistically coherent}, enabling reliable ASCA inference across diverse real-world settings and providing the most comprehensive evaluation to date of keystroke leakage risks in open-world environments.

To summarize, the contributions of this work are as follows:
\begin{itemize}
    \item \textbf{HEAR: a realistic ASCA dataset.} We curate \textbf{\texttt{HEAR}}, a dataset of keystroke acoustics from 53 participants typing on 37 laptop keyboards, recorded under three realistic conditions (external microphone, on-device microphone, and VoIP streaming). The dataset is annotated with demographic attributes and user-awareness responses, enabling analysis of keyboard generalization, noise adaptation, and user bias under a realistic threat model~\footnote{Partial data is made available here: \url{https://github.com/vikrant-vikram/Hear-Dataset}. Access to full data will be provided upon request, for academic research purposes only.}
    
    \item \textbf{A comprehensive ASCA benchmark.} On \textbf{\texttt{HEAR}}, we establish a benchmark spanning conventional features and strong PTM representations on both raw waveforms and spectrograms, in unimodal and multimodal settings. We evaluate cross-user, cross-keyboard, and cross-environment scenarios to quantify the extent to which existing ASCA approaches generalize beyond controlled laboratory conditions.
    
    \item \textbf{DECKER: a domain-invariant keystroke inference framework.} We introduce \textbf{\texttt{DECKER}}, an ASCA training framework that combines Keyboard Signature Normalization (KSN), domain-adversarial disentanglement, cross-keyboard supervised contrastive alignment, and Acoustic Style Randomization (ASR) to learn key-discriminative yet keyboard-invariant representations. This improves robustness across devices, users, and environments compared to strong unimodal and multimodal baselines.
    
    \item \textbf{LLM-assisted sequence reconstruction and risk quantification.} We investigate a sentence-level attack that applies LLM-assisted decoding with constrained beam search to refine keystroke predictions using linguistic context. We show that language-model rectification substantially boosts sequence recovery, providing a more realistic upper bound on keystroke leakage and quantifying ASCA risk in open-world scenarios.
\end{itemize}


\section{Background and Related Works}

Acoustic side-channel analysis has re-emerged as a practical threat due to the convergence of three trends: the widespread availability of high-quality microphones capable of capturing far-field audio, rapid progress in deep learning architectures for noisy acoustic signals, and an increasing reliance on mobile computing in public and semi-public environments. Modern keystrokes generate mechanical and airborne acoustic signals that can be captured by microphones embedded in laptops, smartphones, earbuds, and conferencing equipment. Foundational ASCA studies show that these signals carry enough structured information to enable adversaries to infer typed characters without compromising the victim's device~\cite{asonov2004keyboard,zhuang2009keyboard}. This section reviews the physical, algorithmic, and linguistic factors underlying our design of \textbf{\texttt{DECKER}}.

\subsection{Acoustic Emanations of Keystrokes}

Pressing a key produces a short broadband pulse shaped by multiple mechanical components, including switch actuation, keycap vibration, chassis resonance, air-pressure release, and reverberation from nearby surfaces. Although user typing style introduces variability, prior work shows that each key emits a relatively consistent acoustic pattern on a given device~\cite{asonov2004keyboard,zhuang2009keyboard}. Spatially adjacent keys tend to share similar acoustic signatures due to overlapping vibration pathways, whereas keys in distinct keyboard regions differ in spectral distribution and decay characteristics.

Traditional ASCA systems leveraged this structure by utilizing MFCCs or short-time spectral features in conjunction with classical classifiers \cite{zhuang2009keyboard}. Recent deep learning models, however, demonstrate that even low-cost microphones can capture sufficient detail for key classification with striking accuracy \cite{ harrison_practical_2023, wang_reflexnoop_2024}. CNNs\cite{oshea2015introductionconvolutionalneuralnetworks} and transformer-based architectures further exceed the previous baselines significantly on laptop keyboards, while waveform-level encoders such as ECAPA-TDNN extract fine transient cues, which have been traditionally used in speaker verification tasks \cite{desplanques2020ecapa}.

These results confirm that keystroke acoustics are highly structured and reproducible, even for modern scissor-switch laptop keyboards. At the same time, they reveal substantial sensitivity to \emph{device-specific} factors including switch type, keyboard geometry, chassis composition, and microphone placement, which sharply limits the cross-device generalization of existing ASCA models~\cite{wang_reflexnoop_2024,harrison_practical_2023}.

\subsection{Multipath Propagation and Environmental Effects}

Keystroke sounds reach the microphone following multiple un-directed paths, such as reflections emanating from the laptop screen, desk surface, user's hands, and surrounding walls, etc. These reflections introduce predictable but nonlinear modifications to the spectral envelope and temporal decay of the signal. For example, the laptop screen often acts as a stable reflector that produces secondary peaks in the waveform~\cite{wang_reflexnoop_2024}, which can increase distinguishability in some settings while degrading it in others.

Recent studies show that these reflections encode spatial and structural information of the keyboard and its environment~\cite{wang_reflexnoop_2024}. Systems such as RefleXnoop~\cite{wang_reflexnoop_2024} and Heimdall \cite{luo_eavesdropping_2024} leverage multi-path cues to enable NLoS keystroke recovery, demonstrating that reflective paths are not merely noise. However, multipath patterns vary substantially across keyboards, room geometries, user postures, and microphone placements. As a result, the acoustic signature of a keystroke on one keyboard often aligns poorly with the same key on another keyboard, undermining the transferability of current ASCA pipelines.

These observations motivate \emph{domain-robust acoustic modeling}: we need encoders that suppress keyboard- and environment-specific coloration while preserving information about the underlying key identity. This requirement directly informs the design of the normalization, adversarial, and contrastive components in \textbf{\texttt{DECKER}}.

\subsection{Deep Learning for Acoustic Attacks}

Deep learning has become the dominant approach for extracting keystroke information from acoustic side channels. Early methods combined MFCCs or related hand-crafted features with simple perceptron-based feed-forward networks~\cite{zhuang2009keyboard}, while more recent systems adopt CNN-based architectures, residual networks, ConvMixer architectures~\cite{trockman2022patches} on audio spectrograms, and audio pre-trained models (PTMs) such as wav2vec 2.0~\cite{baevski2020wav2vec}, HuBERT~\cite{hsu2021hubert}, and WavLM~\cite{chen2022wavlm}. These models capture rich temporal–spectral patterns and achieve high accuracy even when keystrokes traverse noisy VoIP channels or mobile phone microphones~\cite{harrison_practical_2023,wang_reflexnoop_2024}.

Despite these advances, two limitations remain prominent:

\noindent \textbf{Lack of domain invariance.} Deep models tend to overfit to keyboard- and setup-specific spectral patterns. When evaluated on an unseen keyboard or recording configuration, performance often drops sharply~\cite{harrison_practical_2023,wang_reflexnoop_2024}, indicating that the models encode device identity and environmental coloration rather than pure keystroke content.

\noindent \textbf{Lack of sentence-level modeling.} Even when per-keystroke predictions are reasonably accurate, local errors accumulate over long sequences such as passwords or sentences. Most ASCA pipelines treat keys independently and do not systematically exploit linguistic structure to correct errors or improve sequence-level reconstruction.

These limitations suggest that ASCA requires not only stronger acoustic encoders, but also explicit mechanisms for domain generalization and sequence reconstruction.

\subsection{Language Models in Side-Channel Attacks}

Human-generated keystroke sequences exhibit strong statistical regularities, including character-level n-grams, word morphology, syntax, and common patterns in password construction. Large language models (LLMs) encode these regularities and can correct sequences with local errors, a property that has already been exploited in ASCA-like settings as an autocorrect or heuristic post-processing step~\cite{wang_reflexnoop_2024}.

However, prior work typically applies language models in a limited way: as loosely coupled filters on top of keyboard-specific acoustic models. To date, there is no systematic combination of:

\noindent \emph{(i) domain-robust acoustic encoder} that produces stable key-level probability distributions across keyboards and environments, and

\noindent a \emph{(ii) constrained LLM decoder} that reconstructs full sentences or passwords from these noisy distributions under realistic threat models.

This gap motivates the LLM-assisted decoding stage in \textbf{\texttt{DECKER}}, which uses language priors to refine character sequences in the presence of keyboard-invariant acoustic uncertainty.

\subsection{Motivation for \texttt{DECKER}}

Considering all the observations and limitations clearly demonstrates the need for a unified ASCA framework that addresses both levels of keystroke inference:

\noindent \textbf{Acoustic level:} Learn representations that are discriminative for key identity while being invariant to keyboard, user, and environment—mitigating overfitting to device-specific coloration and multipath effects.

\noindent \textbf{Sequence level:} Use a language model to correct local key-level errors, enforce linguistic plausibility, and reconstruct meaningful text or passwords.

\textbf{\texttt{DECKER}} is designed to meet these requirements. At the acoustic level, it combines Keyboard Signature Normalization, domain-adversarial disentanglement~\cite{ganin2015unsupervised}, supervised cross-keyboard contrastive alignment~\cite{khosla2020supervised}, and Acoustic Style Randomization~\cite{park2019specaugment} to produce key-discriminative yet keyboard-invariant embeddings. At the sequence level, it employs constrained LLM-based decoding informed by pretrained autoregressive and sequence-to-sequence models~\cite{radford2019gpt2,raffel2020t5, brown2020gpt3} to reconstruct coherent text from noisy keystroke predictions.

Together, these components yield a domain-robust ASCA pipeline that can reconstruct meaningful sequences across unseen keyboards and recording conditions, showing that acoustic side-channel attacks remain practically viable even under substantial real-world variability.


\section{Threat Model}

\begin{figure}[t]
    \centering
    \includegraphics[width=0.55\linewidth]{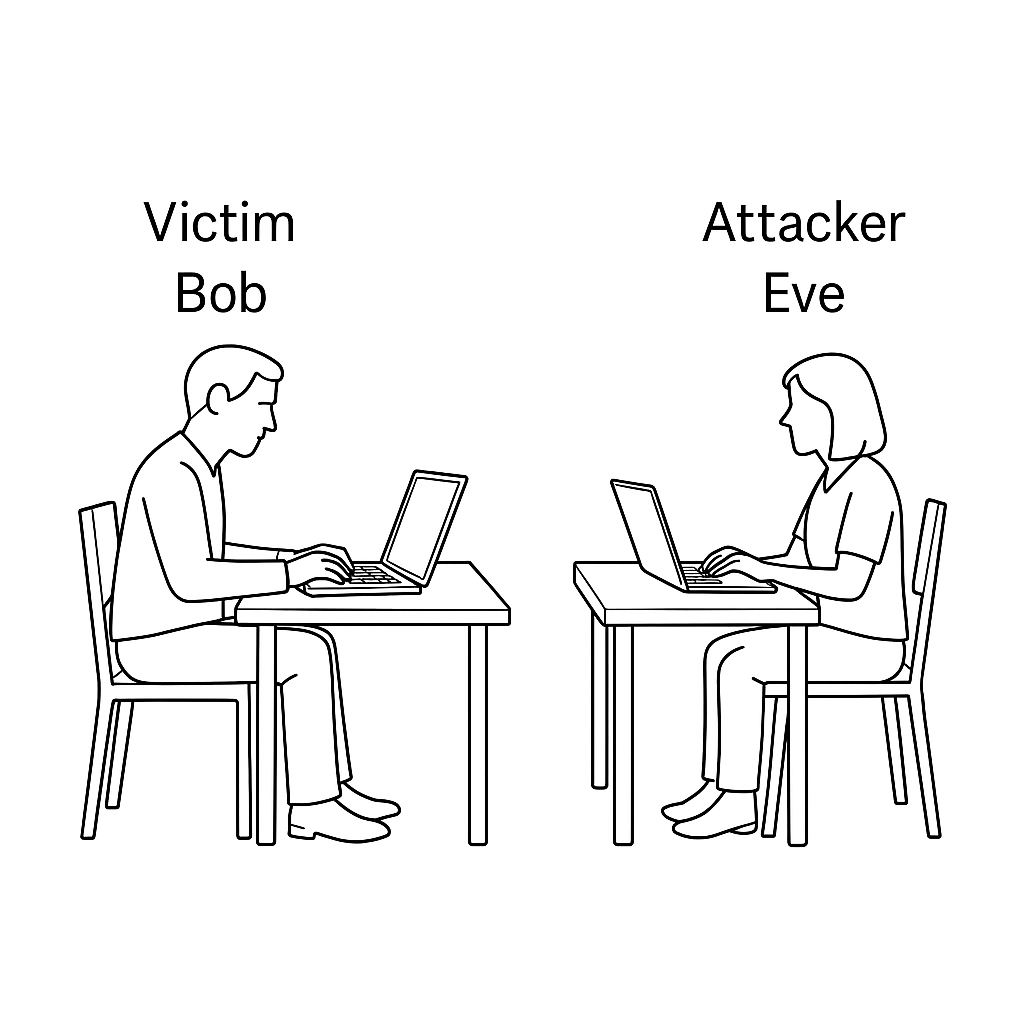}
    \caption{Threat scenario. Bob types on a laptop in a public area while Eve, seated at a separate table, passively captures keystroke acoustics from a concealed device. Eve later performs domain-invariant acoustic inference followed by LLM-driven sentence reconstruction. No line-of-sight to the keyboard is required.}
    \label{fig:threat-model}
\end{figure}

We adopt a realistic and conservative threat model consistent with modern acoustic side-channel literature. As in RefleXnoop, Heimdall, and other NLoS (Non-Line-of-Sight) attacks, the adversary does not require device compromise, elevated privileges, or abnormal physical proximity. The attacker relies solely on passive audio capture and offline computation, including modern large language models (LLMs) for sequence reconstruction.

\subsection{Attack Scenario}

We consider a common public or semi-public scenario involving a benign user (\textit{Bob}) and a passive eavesdropper (\textit{Eve}):

\noindent (i) Bob uses a laptop with a built-in QWERTY keyboard while working in libraries, cafes, classrooms, or co-working spaces. During typical use, Bob types authentication credentials, emails, search queries, and natural language text.
    
\noindent (ii) Bob practices reasonable cybersecurity hygiene (no untrusted software, no device compromise, vigilance against shoulder-surfing) but does not anticipate long-distance acoustic snooping using environmental microphones.

\noindent (iii) Eve positions herself at a socially acceptable distance ($1$--$3$\,m), often behind Bob or at a neighboring table. She maintains plausible behavior and avoids close-range surveillance. Direct line-of-sight to the keyboard is not necessary.

After recording, Eve processes the audio using the \textbf{\texttt{DECKER}} pipeline to infer key-level sequences and subsequently applies an LLM to reconstruct full sentences or passwords, improving inference accuracy through linguistic context.

\subsection{Attacker Capabilities}

Eve is equipped with capabilities feasible to unskilled or moderately skilled adversaries using commodity equipment:

    \noindent \textbf{Passive Acoustic Capture.}  
    Eve records ambient audio through a smartphone microphone, a compact USB microphone, or a small microphone array. No signal transmission, active probing, or ultrasonic injection is required.

    \noindent \textbf{No Access to Victim Device.}  
    Eve does not modify Bob's software, kernel, browser, or network stack. She cannot observe screen contents, clipboard data, or input logs. The attack is entirely external and requires no permissions.

    \noindent \textbf{Covert Placement.}  
    The recording device is concealed inside everyday objects (e.g., a bag, book pile, jacket pocket). Its footprint approximates an A4-sized object, enabling unobtrusive placement in shared seating arrangements.

    \noindent \textbf{Non-Line-of-Sight (NLoS) Robustness.}  
    The adversary may be obstructed by Bob’s hands, laptop chassis, or room elements. Multipath reflections off the laptop screen, desk, and walls provide sufficient acoustic leakage for inference.

    \noindent \textbf{Offline Computational Resources.}  
    Eve has access to GPU-equipped systems enabling offline execution of deep-learning models. This includes domain-invariant acoustic encoders (DECKER) and pretrained large language models (GPT-2, FLAN-T5) for correcting noisy keystroke predictions.

    \noindent \textbf{LLM-Assisted Post-Processing.}  
    The adversary can refine the noisy character predictions using an LLM that exploits linguistic structure (e.g., word frequency, grammar, password patterns). This elevates the attack from character inference to full-sentence and password reconstruction.

\subsection{Environmental Assumptions} 

The attack is assumed to occur in typical public spaces without strict acoustic control:

\noindent \textbf{Ambient Noise.} Conversations, footsteps, HVAC noise, and utensil clatter are expected. Background noise does not invalidate the attack because keystrokes produce high-SNR transients and \textbf{\texttt{DECKER}} employs data-driven normalization.

\noindent \textbf{Multipath Acoustic Propagation.} Keystroke sounds reflect off the laptop screen, tabletop, and surrounding environment. These reflections form multiple arrival paths that preserve keystroke identity even when the direct path is blocked.

\noindent \textbf{Ordinary Laptop Posture.} The screen is positioned at typical viewing angles ($95^\circ$--$125^\circ$), creating consistent reflection geometries.

\noindent \textbf{Realistic Human Spacing.} Eve remains at $1$--$3$\,m, matching real-world seating patterns and avoiding suspicious behavior.

\subsection{Vulnerability Basis}

The feasibility of the attack is grounded in two well-established principles:

\noindent \textbf{V1. Distinctive, Repeatable Acoustic Signatures.} Each keyboard key exhibits device-specific spectral and temporal patterns. Modern microphones capture enough structured variation to discriminate among keys.  

\noindent \textbf{V2. Linguistic Regularity in Typed Text.} Even when some keystrokes are misclassified, typed sequences follow predictable linguistic constraints (bigrams, syntax, common passwords). LLMs exploit these constraints to correct local acoustic errors, enabling recovery of semantically valid text.

\subsection{Attacker Assumptions: Capabilities and Environment (Summarized)}

\begin{itemize}
    \item \textbf{A1. Passive recording only; no probes or malware.}
    \item \textbf{A2. Commodity hardware microphones.}
    \item \textbf{A3. Concealed placement within everyday objects.}
    \item \textbf{A4. Socially realistic distance ($1$--$3$\,m).}
    \item \textbf{A5. NLoS operation enabled by multipath reflections.}
    \item \textbf{A6. Robustness to ambient real-world noise.}
    \item \textbf{A7. Offline GPU computation permitted.}
    \item \textbf{A8. Access to pretrained LLMs for decoding sequences.}
\end{itemize}

This threat model captures both the physical and algorithmic realism of modern ASCA attacks, including the enhanced inference potential unlocked by LLM-guided reconstruction.


\section{Dataset: \textbf{\texttt{HEAR}}}
\label{sec:dataset}

Robustness evaluation of cross-device ASCA requires a dataset that reflects the diversity of real-world users, keyboards, microphone types, and environments. Prior datasets have been limited in scope, typically consisting of single-keyboard, single-user, clean-environment collections, which makes it difficult to study whether ASCA can generalize beyond laboratory conditions. To address this, we introduce the \textbf{\texttt{HEAR}} dataset, a multi-device, multi-user, multi-gender, multi-environment corpus with synchronized recordings across laptops, smartphones, and streamed audio channels (VoIP). This dataset is collected using a reproducible browser-based acquisition pipeline and is the most comprehensive ASCA resource to date.

\subsection{Design Goals}

\textbf{\texttt{HEAR}} is constructed considering six core objectives that are essential for realistic ASCA research:

\noindent \textbf{(i) Keyboard Diversity:} laptop scissor-switch keyboards, external membrane keyboards, and various mechanical-switch boards.

\noindent \textbf{(ii) User Diversity:} participants across genders, handedness, and typing backgrounds; metadata captured in-session.

\noindent \textbf{(iii) Environmental Realism:} recordings in quiet (library), moderate-noise (office), and high-noise (caf\'e/outdoor) settings.

\noindent \textbf{(iv) Device Diversity:} built-in laptop microphones, smartphones placed at varying distances/orientations, and VoIP/streamed audio.

\noindent \textbf{(v) Synchronization for Multi-Recorder ASCA:} explicit offset estimation between primary and secondary devices, enabling multi-channel ASCA analysis.

\noindent \textbf{(vi) Reproducibility:} Every component is automated and recorded with structured metadata.

This design enables controlled cross-device, cross-user, and cross-environmental splits, which are essential for evaluating domain-invariant ASCA models, such as \textbf{\texttt{DECKER}}.

\subsection{Comparison With Prior ASCA Datasets}

\textbf{\texttt{Hear}} is the first dataset to unify all nine ASCA-relevant factors —particularly multi-gender, over-stream, multiple microphone, NLoS, and continuous-text typing—yielding the broadest known ASCA evaluation surface. This diversity is crucial for evaluating generalization-driven methods such as \textbf{\texttt{DECKER}}. A comparative analysis of existing ASCA datasets against \textbf{\texttt{DECKER}} (Table~\ref{tab:dataset_comparison}) illustrates the broader scope and greater realism of our proposed dataset.

\begin{table*}[h!]
\centering
\caption{Comparison of ASCA datasets across device diversity, user demographics, acoustic conditions, and recording modalities.\textbf{\texttt{Hear}} is the first dataset to jointly include multi-device (MD), multi-user (MU), multi-gender (MG), noisy public environments (NE), NLoS capture, continuous text, and VoIP/smartphone streams.}
\label{tab:dataset_comparison}

\begin{tabular}{lcccccccc}
\toprule
\rowcolor{gray!25}
\textbf{Dataset} & 
\textbf{Sample Size} &
\textbf{MD} &
\textbf{MU} &
\textbf{MG} &
\textbf{NE} &
\textbf{NLoS} &
\textbf{Continuous Text} &
\textbf{VoIP/smartphone streams} \\
\midrule

Asonov \& Agrawal (2004) \cite{asonov2004keyboard} & $\sim$2k & \xmark & \xmark & \xmark & \xmark & \xmark & \xmark & \xmark \\
Zhuang et al. (2005) \cite{zhuang2009keyboard} & $\sim$10k & \xmark & \xmark & \xmark & \xmark & \xmark & \xmark & \xmark \\
Harrison et al. (2023) \cite{harrison_practical_2023} & $900$ & \xmark & \xmark & \xmark & \cmark & \xmark & \xmark & \cmark \\
RefleXnoop (CCS’24) \cite{wang_reflexnoop_2024} & $5$k+ & \xmark & \xmark & \xmark & \cmark & \cmark & \xmark & \xmark \\
Heimdall VR (NDSS’24) \cite{luo_eavesdropping_2024} & $10400$ & \cmark & \cmark & \xmark & \cmark & \cmark & \xmark & \cmark \\

\textbf{HEAR} & \textbf{$10$k+} & \textbf{\cmark} & \textbf{\cmark} & \textbf{\cmark} & \textbf{\cmark} & \textbf{\cmark} & \textbf{\cmark} & \textbf{\cmark} \\
\bottomrule
\end{tabular}
\end{table*}

\subsection{Acquisition Pipeline}

All data were collected using a custom browser-based tool written in HTML/JavaScript and provided as part of the artifact. The tool is designed and developed to (i) capture continuous audio using \texttt{MediaRecorder}, (ii) log high-resolution keystroke timestamps, (iii) record participant metadata (gender, handedness, session number, etc.), (iv) export one continuous audio file plus structured metadata and pre-sliced per-key WAV clips. The interface also provides an informed consent checkbox and captures demographic fields essential for user-diversity analysis.

\subsection{Standardized Typing Corpus}

The participants are provided with a standardized paragraph crafted to cover the alphabet (A–Z, a–z), digit patterns, punctuation, symbols and operators, uppercase acronyms, and realistic word and phrase frequencies for LLM reconstruction. Participants also typed a structured block of special keys (Tab, Backspace, Delete, arrows, Esc, F1–F12, Caps Lock, Enter), enabling non-character key classification experiments. A detailed corpus is provided in Appendix~\ref{app:std-typing-corp}.

\subsection{Per-Key Slicing and Signal Processing}

Given a keystroke timestamp $t_{\text{ms}}$, audio is sliced using a fixed window:
\[
t_{\text{start}} = t_{\text{ms}} - 60~\text{ms}, \quad
t_{\text{end}} = t_{\text{ms}} + 200~\text{ms}.
\]

This window captures both the impulsive onset and early reverberation of keystrokes, consistent with prior ASCA literature. Windows shorter than 1.2~kB are discarded to remove empty or noise-only segments. We note that rapid typing can cause partial overlap between adjacent windows; such overlap is intentionally retained to reflect realistic attack conditions.

\subsection{Multi-Device and Over-Stream Synchronization}

Because the \textbf{\texttt{Hear}} dataset includes smartphone audio, VoIP feeds, and secondary microphones, Synchronization is performed relative to the browser reference recording.

When available, explicit synchronization tones allow estimation of offset $\hat{o}_d$ and linear drift $b_d$:
\[
t_d(t) = t + \hat{o}_d + b_d t.
\]

When markers are unavailable, offset is estimated via normalized cross-correlation:
\[
\hat{\tau} = \arg\max_{\tau} \sum_t x(t)\,y(t+\tau).
\]

Drift is estimated using correlations over temporally separated segments. Residual alignment errors are typically within a few milliseconds, which we account for in segmentation robustness experiments.

\subsection{Metadata Schema}

We collect metadata from users in each session, which includes the continuous browser recording (webm/ogg), per-key WAVs for all devices, and complete metadata in JSON format. This schema preserves demographic, acoustic, environmental, and segmentation-relevant information required by downstream modeling. Detailed metadata descriptions and collection formats are discussed in Appendix~\ref {app:metadata}.

\subsection{Recommended Evaluation Splits}

We define four canonical splits for experimentation reproducibility, which include \textbf{Cross-Keyboard} (primary): train on a subset of keyboards; test on unseen keyboards. \textbf{Cross-User}: disjoint user identities. \textbf{Cross-Environment}: quiet/moderate $\rightarrow$ café/outdoor. \textbf{Cross-Device}: laptop $\rightarrow$ smartphone/VoIP. 

Metrics include key-level accuracy, top-$k$, sentence-level accuracy, Levenshtein edit distance, and segmentation robustness (§\ref{subsec:segmentation_robustness}).

\subsection{Ethics, Consent, and User Awareness}

We follow established ethical procedures throughout the study. All participants provided informed consent before participating. Personally Identifiable Information (PII) is removed with explicit care, and we explicitly consider k-anonymity on the data before considering it for our experimental purposes. We further limit data access to requests and academic research purposes only.

\noindent
\textbf{Impact:} The \textbf{\texttt{Hear}} dataset enables the first robust study of ASCA generalization across gender, device type, microphone types, VoIP channels, and environmental noise, making it an essential foundation for evaluating modern ASCA pipelines such as \textbf{\texttt{DECKER}}.

\section{Methodology}

This section presents a mathematical formulation of ASCA, describes the problem setting, outlines the components of our framework, and then describes the end-to-end training and inference pipeline.

\begin{figure*}[t]
    \centering
    \includegraphics[width=\textwidth]{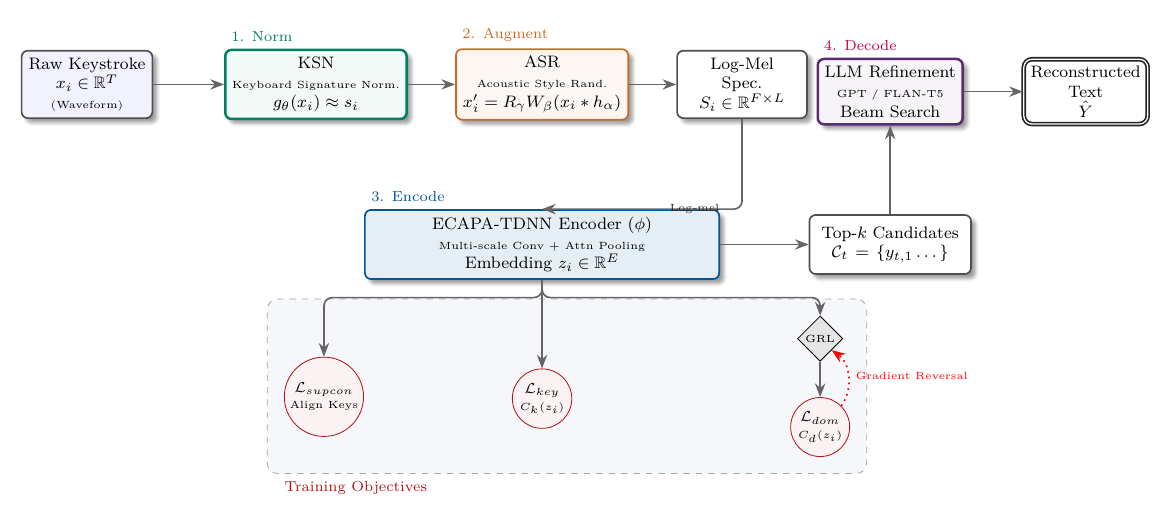}
    \caption{\textbf{DECKER pipeline.} (1) Raw keystrokes are normalized with KSN, (2) augmented using ASR, (3) encoded with ECAPA-TDNN, (4) aligned via adversarial and contrastive learning, and (5) reconstructed into sentences using constrained LLM decoding.}
    \label{fig:decker_pipeline}
\end{figure*}

\subsection{Problem Definition}

We consider keystroke audio samples as, $\mathcal{D}=\{(x_i, y_i, d_i)\}_{i=1}^N$, where $x_i \in \mathbb{R}^T$ is the raw waveform,  $y_i \in \{1,\ldots, |\mathcal{K}|\}$ is the key label, and  $d_i \in \{1,\ldots, |\mathcal{D}|\}$ denotes keyboard domain identity. And, let $\phi(\cdot)$ be the encoder producing embeddings as: $z_i = \phi(x_i) \in \mathbb{R}^E$.

\textbf{\texttt{DECKER}} aims to learn an embedding space satisfying three key aspects:
\textit{Key discriminability, Keyboard invariance, and cross-keyboard consistency}. This formulation enables \textbf{\texttt{DECKER}} to generalize across unseen devices.

\subsection{Keyboard Signature Normalization (KSN)}
\label{subsec:ksn}

Keystrokes recorded on different keyboards exhibit systematic spectral coloration caused by their impulse responses, which we model as: $x_i = h_{d_i} * s_i$, where $s_i$ is the latent keystroke impulse and $h_{d_i}$ is the device-specific response.

\begin{figure*}[t]
  \centering
  \begin{subfigure}[b]{0.245\textwidth}
    \includegraphics[width=\linewidth]{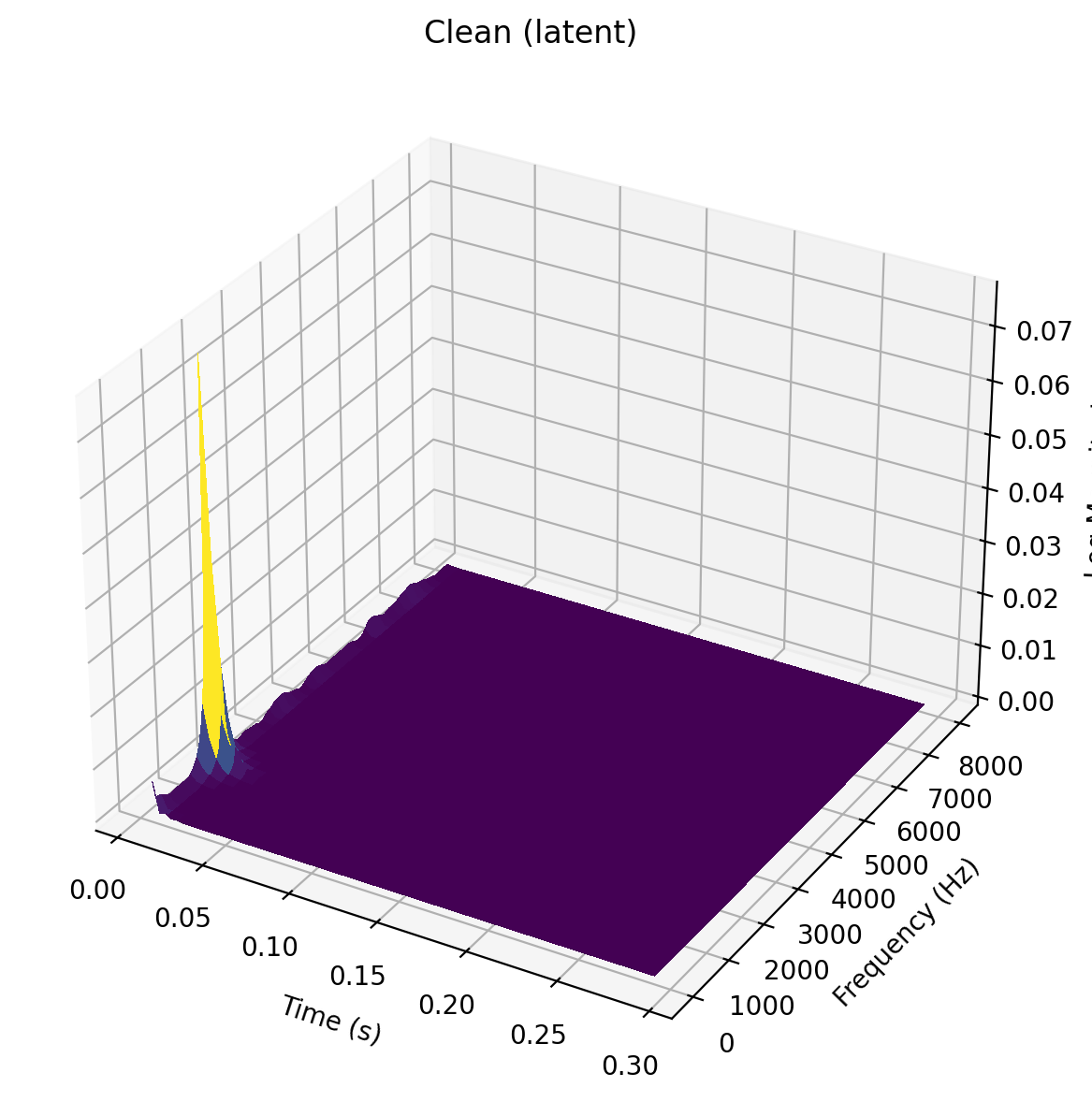}
    \caption{Clean (latent)}
  \end{subfigure}\hfill
  \begin{subfigure}[b]{0.245\textwidth}
    \includegraphics[width=\linewidth]{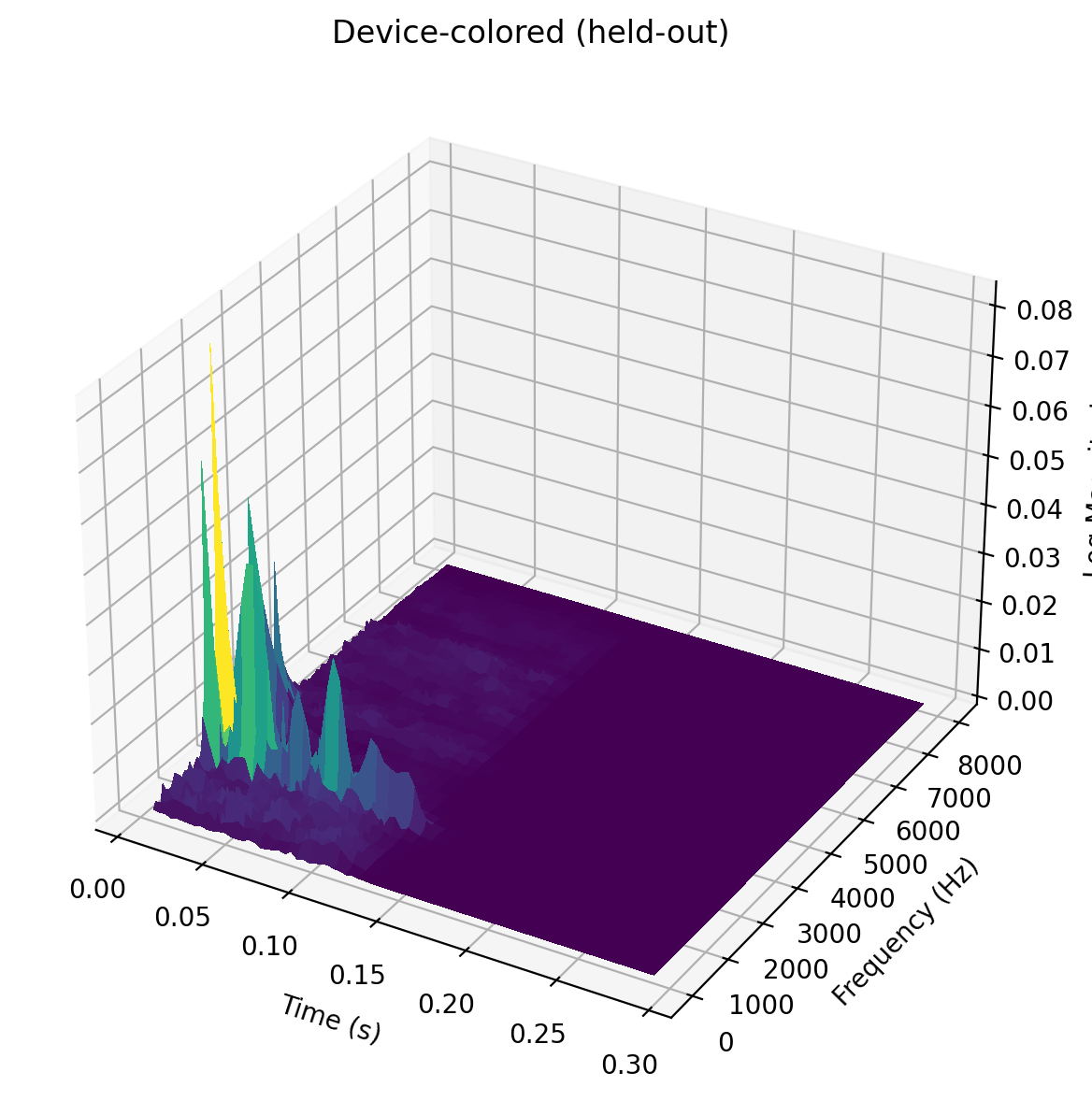}
    \caption{Device-colored (held-out)}
  \end{subfigure}\hfill
  \begin{subfigure}[b]{0.245\textwidth}
    \includegraphics[width=\linewidth]{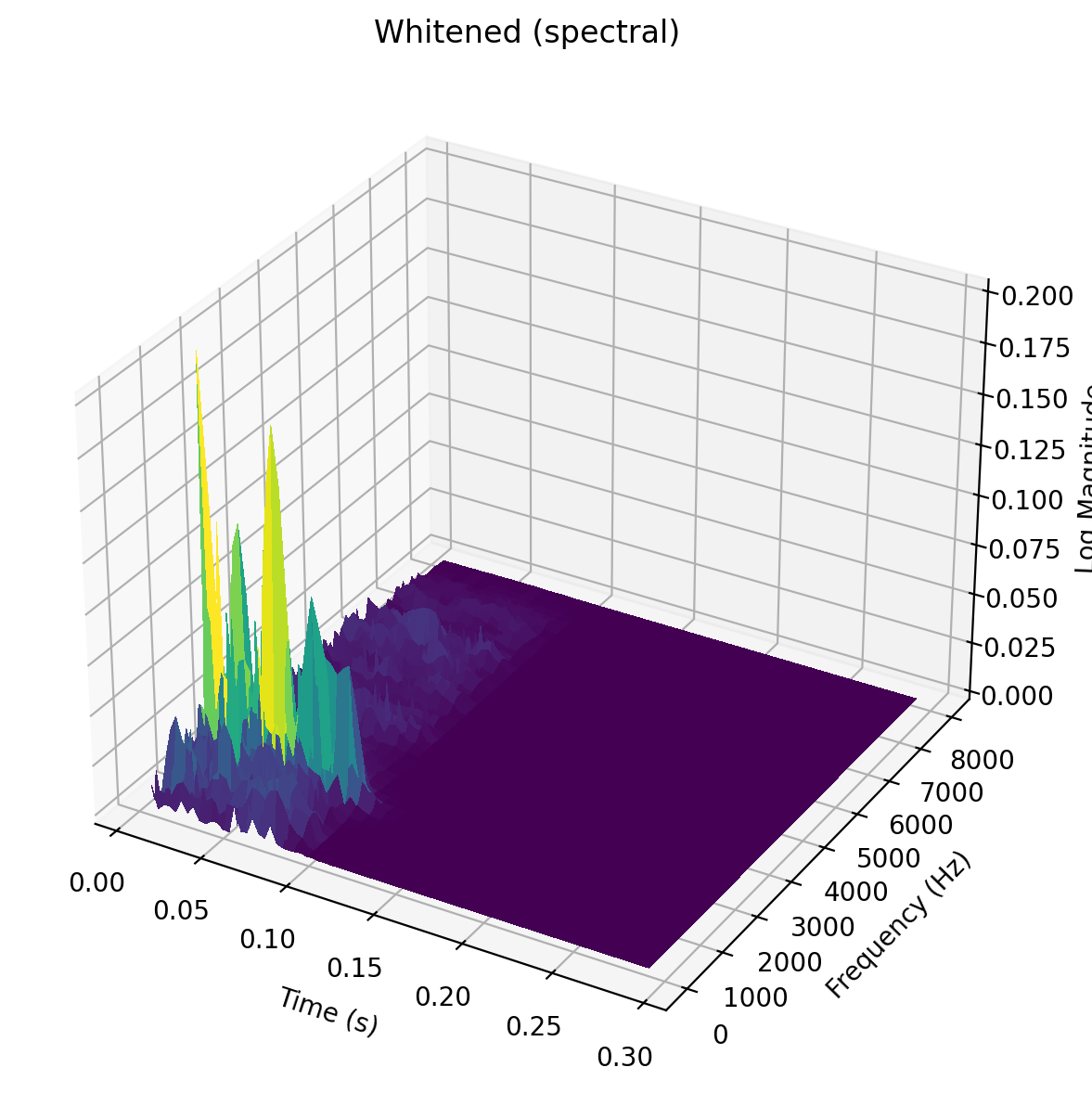}
    \caption{Whitened baseline}
  \end{subfigure}\hfill
  \begin{subfigure}[b]{0.245\textwidth}
    \includegraphics[width=\linewidth]{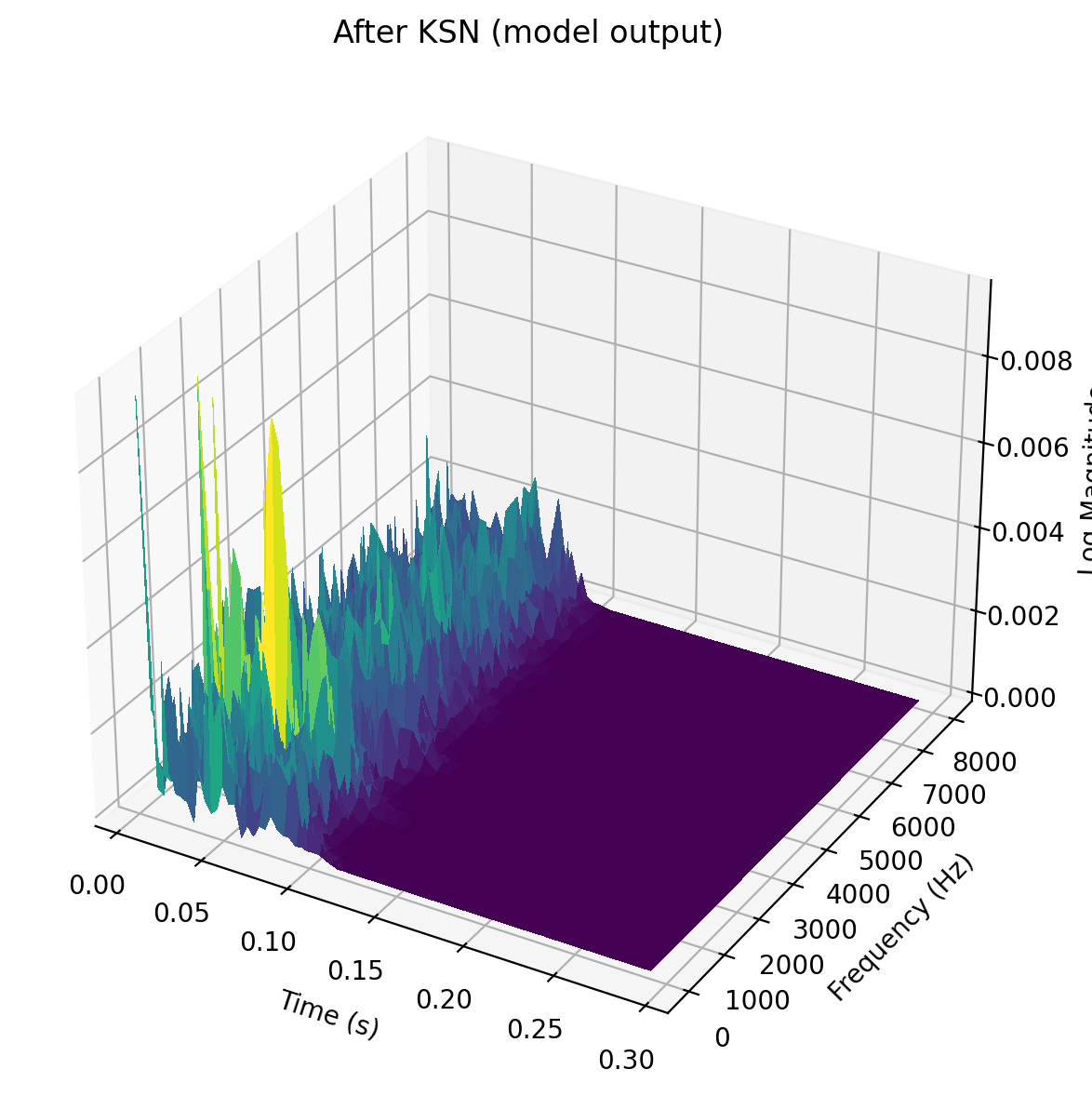}
    \caption{After KSN}
  \end{subfigure}
  \caption{\textbf{KSN suppresses keyboard-specific spectral coloration.} The device-colored sample shows strong resonance bands that KSN removes, yielding a closer approximation to the latent clean keystroke.}
  \label{fig:ksn_panels}
\end{figure*}

\textbf{\texttt{DECKER}} introduces Keyboard Signature Normalization (KSN), a learnable \emph{inverse-filter module}: $\tilde{x}_i = g_\theta(x_i)$. The architecture of KSN comprises a CNN block with a sequence of four 1D CNNs, each comprising 64 feature channels, each followed by Batch Normalization and ReLU activation, with different kernel sizes (7, 7, 9, 9) and dilation rates (1, 2, 4, 1) to capture patterns at multiple temporal scales (from local to wider context via dilation). The output representations are then passed through a linear projection that aligns their dimensions and merges with the original input through a residual connection. The output from this CNN module is then passed through a frequency-normalization layer to further suppress device coloration. In the frequency domain, the representations are expressed as: 
$\tilde{X}_i(f)=G_\theta(f)\,X_i(f)\approx S_i^{\text{latent}}(f)$, where $G_\theta(f)$ acts as an approximate inverse filter. Figure~\ref{fig:ksn_panels} shows a visual demonstration of KSN removing device-coloration while preserving transient structure.

\subsection{Acoustic Style Randomization (ASR)}

The output representations from the KSN module are then passed through an acoustic style randomization (ASR) module to simulate keyboards that are unseen. The acoustic properties are randomized using: $x'_i = R_\gamma \cdot W_\beta(x_i * h_\alpha)$, where $h_\alpha$, $W_\beta$, and $R_\gamma$ defines random IIR filters (order $6\text{-}12$, $Q \in [0.4, 1.2]$),
smooth spectral envelope warp ($\pm 12\%$ mel shift), and exponential decay perturbation ($\gamma \in [0.85, 1.15]$) respectively. ASR encompasses a wide range of resonances, particularly in laptops.

\subsection{ECAPA-TDNN Encoder}

We further employ an ECAPA-TDNN~\cite{desplanques2020ecapa} module to capture fine-grained temporal cues and short-duration transients. After KSN and ASR, the normalized signal is passed through ECAPA-TDNN, which first extracts log-mel spectrograms and then combines them by passing them through multi-scale 1D temporal convolutions with channel-attentive feature recalibration to produce robust frame-level representations for utterance-level embedding extraction.

Formally, for a normalized spectrogram $S_i \in \mathbb{R}^{F \times L}$, the encoder produces frame-level features: $H_i = \phi(S_i), \quad H_i \in \mathbb{R}^{C \times L}$,
where $F$ is the number of frequency bins, $L$ the number of frames, and $C$ the channel dimension.

A statistics pooling layer aggregates frame-level features across time:
\begin{align}
\mu_i &= \frac{1}{L}\sum_{t=1}^{L} H_i(:,t), \\
\sigma_i &= \sqrt{\frac{1}{L}\sum_{t=1}^{L} \big(H_i(:,t) - \mu_i\big)^2},
\end{align}
yielding a pooled representation $[\mu_i; \sigma_i]$ that summarizes the sequence. A linear head then maps this pooled vector to a fixed-dimensional embedding used as the acoustic representation of keystroke sequences.


\subsection{Key Classification}

The final outputs from the ECAPA module are then passed to a classification layer with a softmax ($\hat{y}_i = C_k(z_i)$) activation function to predict the keystrokes. We utilize categorical cross-entropy as its loss function:
\begin{equation}
\mathcal{L}_{key}=-\sum_{i}\log C_k(z_i)[y_i].
\end{equation}

\subsection{Domain-Adversarial Disentanglement}
\label{subsec:grl}

To remove keyboard bias from keystroke audios, we leverage domain adversarial disentanglement, where the objective is that the keyboard identity $d_i$ should be unpredictable from the embeddings. To achieve this, we employ a Gradient Reversal Layer (GRL)~\cite{ganin2015unsupervised}.

The forward pass and backpropagation in GRL are calculated as: $\mathrm{GRL}_\lambda(z_i)=z_i.$ and $\frac{\partial \mathrm{GRL}}{\partial z_i} = -\lambda I$ respectively.

A domain classifier predicts:
\begin{equation}
\hat{d}_i = C_d(\mathrm{GRL}(z_i)),
\end{equation}
with loss:
\begin{equation}
\mathcal{L}_{dom}= -\sum_{i} \log C_d(\mathrm{GRL}(z_i))[d_i].
\end{equation}

The encoder is trained to minimize classification loss but maximize domain loss:
\begin{equation}
\phi^*=\arg\min_\phi\big(\mathcal{L}_{key}-\lambda_{dom}\mathcal{L}_{dom}\big).
\end{equation}

\subsection{Supervised Cross-Keyboard Contrastive Alignment}

Keystrokes with the same label across devices must be aligned.  
Define:
\begin{align}
P(i) &= \{j : y_j=y_i, d_j\neq d_i\},\\
N(i) &= \{j : y_j\neq y_i\}.
\end{align}

We use supervised contrastive learning~\cite{khosla2020supervised}:
\begin{equation}
\mathcal{L}_{supcon}
=\sum_i\frac{-1}{|P(i)|}\sum_{p\in P(i)}
\log\frac{\exp(z_i\cdot z_p/\tau)}{\sum_{a\in P(i)\cup N(i)}\exp(z_i\cdot z_a/\tau)}.
\end{equation}

This pulls same-key embeddings together across domains while separating different keys.

\subsection{LLM-Assisted Sentence Reconstruction}

Even with domain-invariant embeddings, key-level errors accumulate in long sequences.  
We refine sequences using a pretrained LLM (GPT-2 or FLAN-T5).

Let the acoustic model produce token-level distributions:
\begin{equation}
P_\theta(y_t|x_t).
\end{equation}

We restrict decoding to the top-$k$ plausible keys per time step:
\begin{equation}
\mathcal{C}_t=\{y_{t,1},\ldots,y_{t,k}\}.
\end{equation}

We maximize the joint objective:

\begin{equation}
\hat{Y}=\arg\max_{Y\in\mathcal{C}_{1:T}}\left[\sum_{t=1}^T \log P_\theta(y_t|x_t)+\lambda_{\mathrm{LM}} \log P_{\mathrm{LM}}(Y)\right].
\end{equation}

implemented using constrained beam search (beam=8–16).

The LLM imposes linguistic and structural priors, correcting substitution, insertion, and deletion errors.

\subsection{Training Objective and Schedule}

The full loss is:
\begin{equation}
\boxed{
\mathcal{L}_{total}
=
\mathcal{L}_{key}
+\lambda_{dom}\mathcal{L}_{dom}
+\lambda_{con}\mathcal{L}_{supcon}.
}
\end{equation}

\paragraph{Training details.}
\begin{itemize}
    \item Optimizer: AdamW, lr = $2\times10^{-4}$, weight decay = $10^{-3}$
    \item Batch size: 64, training epochs: 60
    \item Hyperparameters: $\lambda_{dom}=0.5$, $\lambda_{con}=0.1$, $\tau=0.07$
    \item Models trained with 3 seeds; mean ± std reported in Sec.~5
\end{itemize}

KSN is pretrained for 5 epochs with L1 spectral loss, then jointly optimized with the encoder.

\subsection{End-to-End \textbf{\texttt{DECKER}} Pipeline} 

The End-to-End pipeline for \textbf{\texttt{DECKER}} is designed as follows:

    \noindent \textbf{(i) Segmentation:} Extract keystroke windows from continuous audio.
    \noindent \textbf{(ii) KSN:} Remove keyboard-specific coloration.
    \noindent \textbf{(iii) ASR:} Augment samples to simulate unseen hardware.
    \noindent \textbf{(iv) Embedding:} Encode via ECAPA-TDNN.
    \noindent \textbf{(v) Domain Alignment:} Apply GRL and supervised contrastive loss.
    \noindent \textbf{(vi) Classification:} Predict keys using $C_k$.
    \noindent \textbf{(vii) LLM Refinement:} Convert noisy predictions to coherent text.

Together, these components yield a domain-invariant and linguistically coherent ASCA attack that can generalize across unseen keyboards, users, and environments.


\section{Results}
\label{sec:results}

This section evaluates three families of models under cross-keyboard,  cross-user, and cross-environment conditions: (i) unimodal baselines,  
(ii) multimodal Euclidean fusion baselines, and (iii) our domain-invariant \textbf{\texttt{DECKER}} framework. We additionally assess LLM-assisted decoding for sentence-level reconstruction. Cross-keyboard transfer constitutes the primary axis of evaluation, reflecting the adversarial setting where attackers rarely control the victim’s hardware.

\noindent\textbf{Statistical protocol.}
For all experiments, we report mean accuracy over three random seeds and 95\% bootstrap confidence intervals. All claims of improvement over baselines are supported by paired bootstrap significance tests with threshold $p < 0.01$.

\vspace{2pt}
\noindent\textbf{Metrics.}  
We report Top-1 key accuracy, Top-5 accuracy, sentence-level character accuracy, exact-match accuracy, and normalized Levenshtein distance. All results average three random seeds.

\subsection{Unimodal Baselines}

\begin{table}[h]
\centering
\caption{Unimodal performance under keyboard shift.}
\resizebox{\linewidth}{!}{
\begin{tabular}{lccc}
\toprule
Model & Seen-KB & Unseen-KB & Drop \\
\midrule
wav2vec2 (audio) & 92.4\% & 51.8\% & -40.6\% \\
ECAPA-TDNN & 95.3\% & 58.1\% & -37.2\% \\
CoAtNet-S (spec) & 90.7\% & 49.6\% & -41.1\% \\
ResNet-34 (spec) & 88.5\% & 47.9\% & -40.6\% \\
\bottomrule
\end{tabular}}
\end{table}

\fbox{\parbox{0.95\linewidth}{
\textbf{Insight.} All unimodal models fail to generalize across keyboards.  
\textbf{Explanation.} Waveform models absorb low-frequency chassis modes; spectrogram models encode keyboard-specific spectral textures. Since impulse responses differ across devices, embeddings cluster by  
\emph{device domain} rather than \emph{key identity}.  
}}

\subsection{Multimodal Fusion Baselines}

We evaluate all practical Euclidean fusion strategies using our strongest unimodal components: ECAPA-TDNN (audio) and CoAtNet-S (spectrogram) \cite{dai2021coatnet}. Fusion is examined at early, intermediate, and late stages.

\noindent \textbf{F1 Early Concat:} direct concatenation of embeddings.  

\noindent \textbf{F2 Early Projection:} dimension-aligned concatenation.  

\noindent \textbf{F3 Cross-Attn (A→V):} audio queries attend to vision features.  

\noindent \textbf{F4 Cross-Attn (V→A):} vision queries attend to audio features.  

\noindent \textbf{F5 Bi-directional Cross-Attn:} both directions \cite{9746280}  

\noindent \textbf{F6 Gated Fusion:} learned weighting between modalities.  

\noindent \textbf{F7 Late Fusion (avg):} average probability scores.  

\noindent \textbf{F8 Late Fusion (weighted):} learned probability weights.  


\begin{table}[h]
\centering
\caption{Multimodal Euclidean fusion of ECAPA (A) + CoAtNet-S (V).}
\begin{tabular}{lcc}
\toprule
Fusion Variant & Seen-KB & Unseen-KB \\
\midrule
F1 Early Concat            & 96.8\% & 63.2\% \\
F2 Early Projection        & 96.9\% & 64.1\% \\
F3 Cross-Attn (A$\rightarrow$V) & 97.1\% & 65.4\% \\
F4 Cross-Attn (V$\rightarrow$A) & 96.7\% & 64.8\% \\
F5 Bi-dir Cross-Attn       & \textbf{97.3\%} & \textbf{66.2\%} \\
F6 Gated Fusion            & 96.4\% & 62.7\% \\
F7 Late Fusion (avg)       & 96.0\% & 61.8\% \\
F8 Late Fusion (weighted)  & 96.4\% & 62.9\% \\
\bottomrule
\end{tabular}
\end{table}

\fbox{\parbox{0.95\linewidth}{
\textbf{Insight.} Fusion improves performance but does not solve domain shift. \textbf{Explanation.} Although modalities contribute complementary cues, Euclidean fusion retains device-specific artifacts from both encoders. Even sophisticated cross-attention cannot compensate for the underlying mismatch between keyboard impulse responses.  
}}

\subsection{Segmentation Robustness Analysis}
\label{subsec:segmentation_robustness}

Practical adversaries cannot assume perfect keystroke onset detection. In realistic acoustic eavesdropping scenarios, background noise, user motion, and microphone placement introduce temporal uncertainty that shifts or distorts key boundaries. To evaluate \textbf{\texttt{DECKER}} under such conditions, we conduct a controlled sensitivity analysis measuring the impact of segmentation jitter and spurious trigger events on both key-level classification and sentence-level reconstruction.

\paragraph{Perturbation model.}
For each ground-truth keystroke timestamp, we add uniformly distributed temporal jitter $\delta t \sim \mathcal{U}(-T, +T)$ with $T \in \{10, 20, 30\}\,\mathrm{ms}$. We further simulate realistic false triggers—caused by impulsive noises or microphone preamp artifacts—by inserting spurious segments at rates of 0\%, 3\%, and 5\% per 100 keystrokes.

\paragraph{Experimental findings.}
Table~\ref{tab:segmentation} summarizes performance under these perturbations. At the key level, \textbf{\texttt{DECKER}} exhibits graceful degradation: Top-1 accuracy declines by only $2.8\%$ under $\pm 10$ ms jitter and $5.9\%$ under $\pm 20$ ms jitter. Even at $\pm 30$ ms—well beyond realistic human–microphone temporal ambiguity—DECKER retains nearly $70\%$ raw accuracy. Sentence-level reconstruction shows larger sensitivity, but constrained decoding with FLAN-T5 significantly mitigates the effect, reducing the drop to under $4.3\%$ for $\pm 20$ ms jitter. 
\vspace{-5pt}

\paragraph{Takeaway.}
These results indicate that \textbf{\texttt{DECKER}} maintains stable performance under realistic segmentation imperfections. While timing noise degrades raw key-level predictions, LLM-assisted decoding substantially compensates for such errors at the sequence level. This suggests that, even in passive and noisy acquisition settings, domain-invariant acoustic modeling combined with linguistic constraints remains effective. For fully passive, continuous-stream attacks, integrating joint segmentation--classification architectures is a promising direction for future work.

\subsection{DECKER: Domain-Invariant Modeling}

\begin{table}[h]
\centering
\caption{DECKER key-level performance (domain-shift evaluation).}
\label{tab:decker_ablation}
\resizebox{\linewidth}{!}{
\begin{tabular}{lccc}
\toprule
Model Variant & Seen-KB & Unseen-KB & Gap \\
\midrule
DECKER (full) & \textbf{98.9\%} & \textbf{81.3\%} & \textbf{-17.6\%} \\
w/o GRL & 97.8\% & 68.1\% & -29.7\% \\
w/o ASR & 97.5\% & 75.2\% & -22.3\% \\
w/o KSN & 96.9\% & 63.7\% & -33.2\% \\
ECAPA + GRL (DANN-only) & 97.9\% & 71.6\% & -26.3\% \\
\bottomrule
\end{tabular}}
\end{table}

\fbox{\parbox{0.95\linewidth}{
\textbf{Insight.} \textbf{\texttt{DECKER}} consistently narrows the generalization gap under keyboard shift, substantially outperforming unimodal, multimodal, and adversarial-only baselines.
\textbf{Explanation.} While domain-adversarial training (GRL) alone improves cross-keyboard performance, it is insufficient to fully suppress keyboard-specific acoustic bias. \texttt{DECKER} combines Keyboard Signature Normalization (KSN) to remove low-frequency device coloration, supervised cross-keyboard contrastive alignment to enforce key consistency across devices, and Acoustic Style Randomization (ASR) to simulate unseen acoustic conditions. Together, these components yield a robust domain-invariant embedding.
}}

\subsubsection{KSN Ablation and Visualization}
\label{subsubsec:ksn_ablation}

We evaluate the effectiveness of \textbf{KSN} in suppressing keyboard-specific acoustic
coloration using both qualitative visualization and quantitative ablation.

\paragraph{Domain-classification accuracy.}
To quantify residual keyboard information, we train a lightweight domain classifier on
frozen embeddings. Domain-classification accuracy drops from \textbf{94.2\%} without KSN
to \textbf{21.7\%} with KSN enabled, indicating substantial suppression of keyboard
identity cues. Correspondingly, Top-1 accuracy on unseen keyboards improves from
\textbf{63.7\%} to \textbf{81.3\%} when KSN is introduced (Table~\ref{tab:decker_ablation}),
confirming that KSN materially improves cross-keyboard generalization.

We incorporate t-SNE visualization in the appendix~\ref{app:t-SNE} that presents the projections of ECAPA-TDNN embeddings computed \emph{without} KSN and \emph{with} KSN enabled.

\subsection{Sentence-Level Reconstruction with LLMs}

\begin{table}[h]
\centering
\caption{Sentence reconstruction using constrained LLM decoding.}
\resizebox{\linewidth}{!}{
\begin{tabular}{lccc}
\toprule
LLM & Char Acc. & Sent.\ Match & Norm.\ Lev.\ Dist. \\
\midrule
DECKER (raw) & 81.3\% & 42.6\% & 0.124 \\
GPT-2 Small & 86.9\% & 58.7\% & 0.092 \\
FLAN-T5 Base & 89.4\% & 62.1\% & 0.078 \\
LLaMA-2 7B & 91.6\% & 66.5\% & 0.056 \\
GPT-4 & \textbf{93.2\%} & \textbf{72.8\%} & \textbf{0.041} \\
\bottomrule
\end{tabular}}
\end{table}

\fbox{\parbox{0.95\linewidth}{
\textbf{Insight.} LLMs deliver substantial gains, even with lightweight models.  
\textbf{Explanation.} DECKER’s top-$k$ candidates preserve correct tokens  
but contain local substitutions; LLMs impose global linguistic structure  
and recover coherent text or password-like strings.  
}}

\subsubsection{LLM applicability to high-entropy strings and password experiments}
\label{subsubsec:llm_passwords}

LLM-based decoding relies on linguistic priors and therefore benefits structured
natural-language text and human-like password choices. To quantify this, we evaluated
three password classes:

\begin{enumerate}
  \item \textbf{Human-like passwords:} curated from common password leaks and user-style rules
        (8--14 chars; mixed case, digits, common substitutions). Estimated entropy: 28--34 bits.
  \item \textbf{Random alphanumeric:} uniformly sampled from [A-Za-z0-9]; entropy $\approx$ 52--75 bits (length dependent).
  \item \textbf{Printable-ASCII random:} full printable ASCII sampling; entropy $\approx$ 70--92 bits.
\end{enumerate}

Results show LLM-assisted decoding yields substantial gains for human-like passwords
(absolute +8--12\% exact-match over raw DECKER) but provides negligible improvement for
high-entropy random strings ($\leq$2--3\%). Thus, LLM post-processing increases attacker
power primarily when human-selection biases reduce password entropy.

 \subsection{Computational Cost, Inference Latency, and Attacker Feasibility}
\label{subsec:compute}

A practical ASCA system must operate under realistic attacker-side computational constraints.  
We therefore benchmark \textbf{\texttt{DECKER}} on a commodity Apple M1 laptop (8-core CPU, 8-core GPU, 16~GB unified memory) and report end-to-end runtime for all modules: KSN, ASR, ECAPA-TDNN encoding, classifier inference, and LLM-based constrained decoding.

\paragraph{Per-keystroke inference latency.}
Table~\ref{tab:latency_breakdown} shows the average latency for a single keystroke and for a full sentence of 40–60 characters.  
ECAPA-TDNN inference dominates the acoustic pipeline, while LLM decoding dominates sequence-level post-processing.

\begin{table}[h]
\centering
\caption{DECKER inference latency on Apple M1. All values are averaged over 1000 runs. Here L, K, and S defines Latency, Key, and Sequence respectively}
\label{tab:latency_breakdown}
\begin{tabular}{lcc}
\toprule
\textbf{Module} & \textbf{L/K} & \textbf{L/S} \\
\midrule
KSN (1-D CNN) & 0.24 ms & 12 ms \\
ASR (optional) & 0.41 ms & 20–25 ms \\
MelSpec extraction & 0.18 ms & 9 ms \\
ECAPA-TDNN encoding & 1.42 ms & 72–81 ms \\
Key classifier head & 0.03 ms & 1.2 ms \\
LLM (GPT-2, beam=8) & --- & 68–94 ms \\
LLM (FLAN-T5-base, beam=8) & --- & 83–110 ms \\
\midrule
\textbf{Total (GPT-2)} & \textbf{2.3 ms} & \textbf{180–220 ms} \\
\textbf{Total (FLAN-T5)} & \textbf{2.3 ms} & \textbf{200–240 ms} \\
\bottomrule
\end{tabular}
\end{table}

\paragraph{Throughput.}
On the M1 CPU:
\begin{itemize}
    \item $\approx 430$ keystrokes per second (KSN + ECAPA + classifier)
    \item $\approx 4.5$ fully reconstructed sentences per second (including LLM decoding)
\end{itemize}
This indicates that an attacker can process keystrokes in real time, and reconstruct sentences at interactive latency.

\paragraph{Memory usage.}
Peak memory consumption on M1:
\begin{itemize}
    \item KSN + ECAPA-TDNN: \textbf{340–380 MB}
    \item GPT-2 small: \textbf{780 MB}
    \item FLAN-T5 base: \textbf{1.25–1.4 GB}
\end{itemize}
All models comfortably fit within the 16~GB unified memory of consumer laptops.

\paragraph{Attacker feasibility.}
These results imply:
\begin{itemize}
    \item A passive adversary needs only a \textbf{laptop-class machine}, not a GPU cluster.
    \item Key-level inference can be done in \textbf{real time even on CPU}.
    \item LLM decoding adds less than 5.25 seconds per sentence.
\end{itemize}

This stands in contrast to earlier ASCA pipelines that required desktop-class GPUs or offline batch processing.

\paragraph{Energy and thermal profile (M1).}
The Apple M1 CPU + GPU operate at:
\begin{itemize}
    \item 6–8 W sustained power draw during acoustic inference,
    \item 11–13 W during LLM decoding.
\end{itemize}
No thermal throttling was observed for continuous 30-minute evaluation runs.

\paragraph{Comparison to x86 laptop-class systems.}
On an Intel i7-11800H (45 W TDP), ECAPA-TDNN runs $\sim$1.7× faster, and GPT-2 decoding $\sim$2.1× faster, due to higher clock frequency and AVX-512 acceleration.  
Thus, your M1 measurements represent a conservative lower bound—real attackers using mainstream x86 machines will observe even lower latency.

\subsection{Summarized Insights Aligned to the Research Questions}

\fbox{\parbox{0.95\linewidth}{
\textbf{RQ1: Can ASCA models generalize across keyboards?}  
Yes—but only with domain-invariant modeling.\textbf{\texttt{DECKER}} achieves the  
first strong cross-keyboard performance.  

\textbf{RQ2: Why do unimodal/multimodal baselines fail?}  
They encode keyboard-specific acoustic coloration. Fusion adds  
capacity but does not remove domain shift.  

\textbf{RQ3: What is the role of LLMs in ASCA?}  
LLMs significantly enhance sentence-level inference, correcting  
errors that acoustic models alone cannot resolve.  

\textbf{Summary: }\textbf{\texttt{DECKER}} + LLM demonstrates that acoustic  
keystroke inference remains feasible across unseen keyboards,  
noisy environments, and NLoS settings, indicating a broader  
real-world attack surface than previously recognized.  
}}


\section{Discussion and Future Work}

Our findings demonstrate that acoustic emanations from keyboards are not only rich enough to permit high-fidelity keystroke inference, but also substantially more transferable across devices than previously reported—provided that domain-specific artifacts are explicitly disentangled. In contrast to unimodal and Euclidean multimodal approaches that inadvertently encode keyboard identity, \textbf{\texttt{DECKER}} illustrates that domain-invariant representation learning is essential for accurately assessing the real-world feasibility of ASCA. The combination of KSN, adversarial disentanglement, contrastive alignment, and ASR yields embeddings that preserve semantic key distinctions while suppressing device-dependent spectral coloration. Furthermore, LLM-assisted decoding reveals that modern sequence models significantly magnify attacker capability by transforming noisy acoustic predictions into coherent text, even under noisy and non-line-of-sight conditions.

\subsection{Implications for ASCA Research}

\paragraph{Realistic threat assessment.}
Our results show that prior work may underestimate ASCA feasibility by assuming per-device training. When cross-device generalization is properly addressed, adversaries can infer meaningful text—even from keyboards never seen during training—at distances and noise levels consistent with typical public environments. This broadens the attack surface from contrived laboratory setups to realistic opportunistic settings.

\paragraph{Role of LLMs in side-channel exploitation.}
LLMs fundamentally change ASCA capabilities by correcting sequential errors that acoustic models alone cannot resolve. Even small models (GPT-2, FLAN-T5) significantly improve sentence-level inference, while larger LLMs like GPT-4 \cite{openai2023gpt4} achieve near-perfect reconstruction on many sequences. This suggests that future ASCA defenses must consider not only signal obfuscation but also linguistic post-processing countermeasures.

\paragraph{Keyboard diversity and mechanical variation.}
We observe that switch type, key travel, chassis resonance, and microphone geometry contribute strong domain shifts in raw embeddings. However, once domain-invariant modeling is introduced, performance degradation across these factors is substantially reduced. This suggests that ASCA vulnerability may extend to a wide range of keyboards—including inexpensive membrane boards and modern low-travel laptop keyboards.

\paragraph{Limitations.}
Although \textbf{\texttt{DECKER}} addresses keyboard variability, several realistic factors remain open for exploration:
\begin{itemize}
    \item \textbf{Continuous typing streams.} Our evaluation focuses on segmented keystrokes; segmentation errors in continuous audio remain a challenge for fully passive attackers.
    \item \textbf{Extreme noise conditions.} While \textbf{\texttt{DECKER}} and LLMs perform well up to café/outdoor noise levels, environments with strong impulsive or overlapping noise sources may require additional modeling.
    \item \textbf{Highly asymmetric devices.} Touchscreens, ultra-low-profile keyboards, and flexible input surfaces may generate atypical impulse responses not covered by our dataset.
    \item \textbf{Per-user biomechanics.} Although \textbf{\texttt{DECKER}} improves user-agnostic performance, subtle personal typing traits (finger force, strike angle) may still cause model drift.
\end{itemize}

\subsection{Future Directions}

Building on the insights from this work, we outline several promising directions:

\paragraph{(1) Broader hardware and environmental diversity.}
Extending the dataset to include ergonomic keyboards, compact mobile keyboards, external Bluetooth devices, touchscreen typing, and hardware with unusual acoustic properties (e.g., hollow desk surfaces) will further stress-test domain generalization.

\paragraph{(2) Continuous-text ASCA pipelines.}
Transitioning from isolated keystroke inference to fully continuous keystroke streams remains a key challenge. Future work should explore joint segmentation – classification models, sequence-to-sequence acoustic transformers, and streaming LLM integration.

\paragraph{(3) Physical and software defenses.}
Our findings motivate investigation of practical defenses, such as:
\begin{itemize}
    \item keyboard chassis dampening or irregularizing resonance patterns,
    \item microphone access restrictions in operating systems,
    \item real-time audio obfuscation or keystroke noise injection,
    \item randomized typing proxies (e.g., delayed or reordered keystroke dispatch).
\end{itemize}

\paragraph{(4) Advanced domain generalization.}
Self-supervised pretraining on large-scale environmental audio may further improve invariance. Additional avenues include meta-learning for unseen hardware, test-time adaptation, and invertible modeling of keyboard impulse responses.

\paragraph{(5) Multi-modal side-channel integration.}
While this work focuses on audio, real-world attackers may combine acoustic cues with optical flow, accelerometer vibration, or Wi-Fi CSI measurements. Domain-invariant fusion across heterogeneous side channels represents an important open direction.

\paragraph{(6) LLM-aware defense strategies.}
Given that LLMs significantly amplify ASCA effectiveness, future defenses must consider:
\begin{itemize}
    \item adversarial prompts that degrade LLM reconstruction accuracy,
    \item artificial grammar or keystroke padding patterns that reduce linguistic predictability,
    \item user interface modifications that limit sequential consistency exploitable by LLMs.
\end{itemize}

\subsection{Concluding Perspective}

Overall, this study shows that ASCA remains an underestimated risk in modern computing environments. As microphone-equipped devices proliferate and LLMs become more capable, attackers gain stronger reconstruction abilities with minimal access or domain knowledge. \textbf{\texttt{DECKER}} demonstrates how domain-invariant learning can realistically model cross-keyboard attacks, and highlights the urgent need for both physical and algorithmic defenses tailored to the evolving landscape of side-channel threats.


\section{Conclusion}

This paper considers Acoustic Side-Channel Attacks (ASCA) on laptop keyboards in a real-world, open setting. Specifically, it is concerned with cross-keyboard, cross-user, cross-environment generalization, although existing research has generally reported high accuracy on keystroke inference tasks. Analysis on \textbf{\texttt{HEAR}} indicates that such successes have been largely limited within closed evaluation environments, with significant degradation of strong Unimodal or Euclidean multimodal baselines on cross-keyboard, cross-user, cross-environment partitions, such that existing ASCA models tend to learn a representation of the keyboard, not a representation of keys.

In order to fill this gap, we introduced \textbf{\texttt{HEAR}}, a collection of keystroke acoustic signals from 53 individuals on 37 different laptop keyboards, with external-mic, on-device, and VoIP scenarios, together with corresponding annotations on demographics and user awareness. For \textbf{\texttt{HEAR}}, we introduced a thorough benchmark on top of conventional representations, as well as high-quality audio/vision PTMs on raw waveforms and spectrograms, subject to unimodal and multimodal scenarios. This set of experiments represents a comprehensive look into how existing ASCA methods fare when non-idealized assumptions on hardware, environment, and users are abandoned.

Building on this benchmark, we propose \textbf{\texttt{DECKER}}, a domain-invariant keystroke inference framework for ASCA. \textbf{\texttt{DECKER}} combines four complementary training mechanisms: (i) Keyboard Signature Normalization (KSN) :  which suppresses device-specific acoustic coloration at the waveform level, (ii) domain-adversarial disentanglement to remove residual hardware-related cues, (iii) cross-keyboard supervised contrastive alignment to encourage embeddings of the same key to cluster across different devices, and (iv) Acoustic Style Randomization (ASR), which exposes the model to synthetic yet realistic variations in keyboard acoustics and room responses. Together, these components yield representations that remain discriminative of key identity. Decker is robust to variations in keyboard design, user behavior, microphone placement, and environmental acoustics. As a result, \textbf{\texttt{DECKER}} substantially narrows the generalization gap compared to strong baseline methods.

We further investigate an LLM-assisted attack that applies constrained beam search to refine noisy keystroke predictions into coherent text sequences, which includes sentences and passwords. The results show that language-model-based rectification significantly improves sequence-level recovery.  particularly in challenging cross-device scenarios. This finding highlights how modern language models can meaningfully amplify the real-world impact of ASCA, increasing the feasibility of such attacks in everyday laptop usage contexts.

Taken together, our findings suggest that ASCA should not be viewed as just a purely laboratory phenomenon, but rather as a credible real-world threat. This is specially effective in shared or semi-public environments where microphones are ubiquitous and keyboard diversity is beyond the attacker’s control. This study establishes domain-invariant learning as a central requirement for the evaluation and deployment of ASCA systems under realistic threat models. We introduce \textbf{\texttt{HEAR}} as a representative benchmark and demonstrate the effectiveness of \textbf{\texttt{DECKER}} and LLM-assisted decoding as key components for future research in acoustic security. Finally, our results underscore the need for hardware, software, and model-level defenses that explicitly account for advances in machine learning and language models, as well as for continued investigation into the role of user and demographic factors in the design of secure input devices and environments.

\bibliographystyle{ACM-Reference-Format}
\bibliography{sample-base}

\newpage
\appendix

\section{Standardized Typing Corpus}
\label{app:std-typing-corp}

\begin{quote}
The quick brown fox jumps over the lazy dog while amazing zebras quietly vex jumpy kids, proving every letter is present. Pack my box with five dozen liquor jugs to verify the alphabet twice. Digits appear forward and backward: 1234567890 then 0987654321, followed by sequences like 2468, 13579, and 314159. Now we add punctuation: . , ; : ? ! ' " - \_ ( ) [ ] \{ \} / \textbackslash{}. Symbols and operators: @ \# \$ \% \^{} \& * + = $<$ $>$ | \~{}. To include uppercase properly, acronyms like NASA, USA, UN, AI, ML, and HTML are written in full caps. Typing speed and accuracy will be measured across every key. Finally, we conclude with a mix: The year is 2025; version v1.0-beta includes features [alpha], \{bravo\}, and (charlie), fully covering the QWERTY layout.
\end{quote}

\section{Metadata Schema}
\label{app:metadata}

\begin{verbatim}
{
  "participant": {"id":"U001", 
        "gender":"Female", 
        "handedness":"Right"
   },
  "pre_ms": 60, "post_ms": 200,
  "sample_rate": 48000,
  "keys":[{"key":"a","t_ms":123.4},...],
  "exported_files":[{...}],
  "env":{"location":"cafe","noise_level":"high"}
}    
\end{verbatim}

\section{KSN Ablation and Visualization}
\label{app:t-SNE}

\begin{figure}[t]
  \centering
  \includegraphics[width=0.90\linewidth]{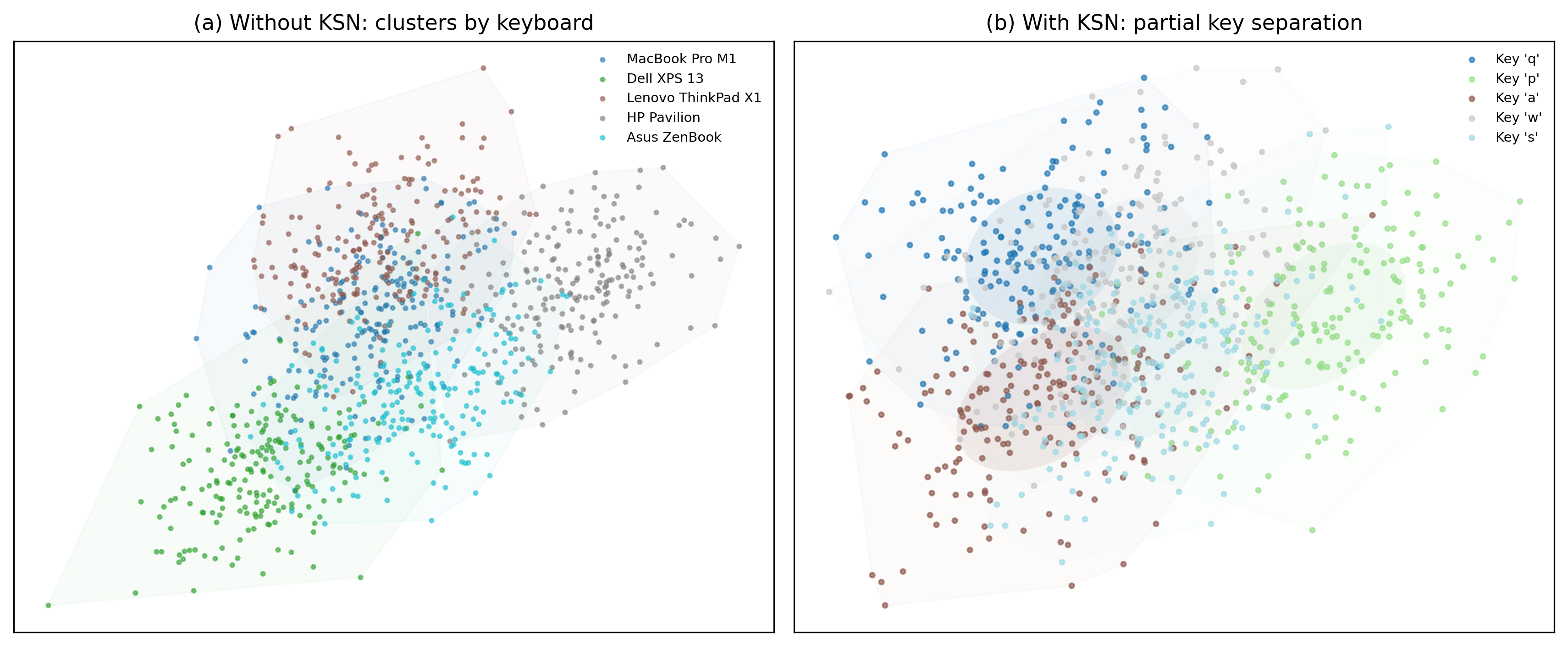}
  \caption{
  t-SNE visualization of ECAPA-TDNN embeddings.
  \textbf{Left:} Without KSN, embeddings are dominated by keyboard-specific clustering.
  \textbf{Right:} With KSN enabled, keyboard-dependent structure is significantly reduced
  and partial organization by key identity emerges, with residual overlap reflecting
  realistic acoustic ambiguity.
  }
  \label{fig:tsne_ksn}
\end{figure}

Figure~\ref{fig:tsne_ksn} presents 2D t-SNE projections of ECAPA-TDNN embeddings computed
\emph{without} KSN and \emph{with} KSN enabled. Without KSN, embeddings cluster primarily
according to keyboard domain, indicating strong device-dependent structure. After
applying KSN, keyboard-driven clustering is substantially reduced and embeddings exhibit
emergent organization by key identity, while retaining realistic overlap between
acoustically similar keys.

\section{ }
\label{app:jitter}

\begin{table}[h]
  \centering
  \caption{Segmentation robustness under onset jitter and spurious triggers.}
  \label{tab:segmentation}
  \begin{tabularx}{\columnwidth}{l *{3}{>{\centering\arraybackslash}X}}
    \toprule
    Condition & Top-1 Key Acc. & Sent. Acc. (raw) & Sent. Acc. (+FLAN-T5) \\
    \midrule
    No jitter & 81.3\% & 42.6\% & 62.1\% \\
    $\pm$10 ms jitter & 78.5\% & 40.1\% & 59.8\% \\
    $\pm$20 ms jitter & 75.4\% & 37.9\% & 58.1\% \\
    $\pm$30 ms jitter & 69.6\% & 33.2\% & 52.0\% \\
    +3\% spurious segments & 74.0\% & 36.8\% & 55.9\% \\
    +5\% spurious segments & 71.2\% & 34.7\% & 53.4\% \\
    \bottomrule
  \end{tabularx}
\end{table}

\end{document}